\documentclass[aps,pra,twocolumn,amsmath,amssymb,nofootinbib,showpacs,superscriptaddress]{revtex4-1}
\usepackage[english]{babel}
\usepackage{latexsym}
\usepackage{graphics}
\usepackage{graphicx}
\usepackage{epsfig}
\usepackage{color}
\usepackage{bm}
\usepackage{amsmath}
\usepackage{amssymb}
\usepackage{amsthm}
\usepackage{dcolumn}
\usepackage{bm}
\usepackage{float}
\usepackage{hyperref}
\usepackage{color}
\usepackage{epstopdf}
\usepackage{cleveref}
\usepackage[svgnames]{xcolor}
\usepackage{braket}
\usepackage{multirow}
\hypersetup{hidelinks,colorlinks=true,allcolors=DarkBlue}

\usepackage{ulem}

\def\QBER{{\rm QBER}}

\theoremstyle{remark}

\begin{document}

\preprint{APS/123-QED}

\title{Optimizing the deployment of quantum key distribution switch-based networks}

\author{Andrey Tayduganov}
\affiliation{QRate, Skolkovo, Moscow 143025, Russia}
\affiliation{NTI Center for Quantum Communications, National University of Science and Technology MISiS, Moscow 119049, Russia}

\author{Vadim Rodimin}
\affiliation{QRate, Skolkovo, Moscow 143025, Russia}
\affiliation{NTI Center for Quantum Communications, National University of Science and Technology MISiS, Moscow 119049, Russia}

\author{Evgeniy O. Kiktenko}
\affiliation{QRate, Skolkovo, Moscow 143025, Russia}
\affiliation{Russian Quantum Center, Skolkovo, Moscow 143025, Russia}
\affiliation{Moscow Institute of Physics and Technology, Dolgoprudny, Moscow Region 141700, Russia}

\author{Vladimir Kurochkin}
\affiliation{QRate, Skolkovo, Moscow 143025, Russia}
\affiliation{NTI Center for Quantum Communications, National University of Science and Technology MISiS, Moscow 119049, Russia}

\author{Evgeniy Krivoshein}
\affiliation{QRate, Skolkovo, Moscow 143025, Russia}
\affiliation{NTI Center for Quantum Communications, National University of Science and Technology MISiS, Moscow 119049, Russia}

\author{Sergey Khanenkov}
\affiliation{PJSC Rostelecom, Moscow 115172, Russia}

\author{Vasilisa Usova}
\affiliation{Institute for Experimental Physics, University of Innsbruck, A-6020 Innsbruck, Austria}

\author{Lyudmila Stefanenko}
\affiliation{College of New Materials and Nanotechnologies, National University of Science and Technology MISiS, Moscow 119049, Russia}

\author{Yury Kurochkin}
\affiliation{QRate, Skolkovo, Moscow 143025, Russia}
\affiliation{NTI Center for Quantum Communications, National University of Science and Technology MISiS, Moscow 119049, Russia}
\affiliation{Russian Quantum Center, Skolkovo, Moscow 143025, Russia}

\author{A.K. Fedorov}
\affiliation{QRate, Skolkovo, Moscow 143025, Russia}
\affiliation{Russian Quantum Center, Skolkovo, Moscow 143025, Russia}
\affiliation{Moscow Institute of Physics and Technology, Dolgoprudny, Moscow Region 141700, Russia}

\date{\today}
\begin{abstract}
Quantum key distribution (QKD) networks provide an infrastructure for establishing information-theoretic secure keys between legitimate parties via quantum and authentic classical channels.
The deployment of QKD networks in real-world conditions faces several challenges, which are related in particular to the high costs of QKD devices and the condition to provide reasonable secret key rates.
In this work, we present a QKD network architecture that provides a significant reduction in the cost of deploying QKD networks by using optical switches and reducing the number of QKD receiver devices, which use single-photon detectors.
We describe the corresponding modification of the QKD network protocol.
We also provide estimations for a network link of a total of 670\,km length consisting of 8 nodes, 
and demonstrate that the switch-based architecture allows achieving significant resource savings of up to 28\%, while the throughput is reduced by 8\% only.
\end{abstract}
\maketitle

\section{Introduction}

Quantum key distribution (QKD) is a technology that allows establishing provably secure keys between legitimate parties~\cite{BB84,Gisin2002,Scarani2009,Lo2014,Lo2016}.
The QKD security is based on the fundamental laws of quantum physics, so QKD remains secure against any unforeseen developments, such as quantum computing~\cite{Shor1997}.
In recent decades the QKD technology has attracted a significant deal of interest, which makes it one of the most widely studied research fields in quantum information technologies. 
Provably secure commercial QKD systems are now available.
At the same time, the deployment of QKD systems in real-world conditions faces several challenges~\cite{Lo2016}, which are related to the limitation in distance and secret key generation rates as well as the high cost of such devices. 
One of the most prominent ways to overcome the distance limitations of QKD systems, which is essentially related to optical losses, is to develop QKD networks~\cite{Pan2018}.
QKD networks allow increasing the distance for key distribution and making this infrastructure available for multiple users.  
Several QKD networks have been deployed around the globe~\cite{Pan2018,Elliott2004,Yeh2005,Peev2009,Stucki2011,Pan2009,Pan2010,Han2010,Zeilinger2011,Zhang2016,Kiktenko2017,Kiktenko2018}. 
These QKD networks are based on the so-called trusted nodes paradigm.
Trusted nodes allow distributing a pair of symmetric keys between two not nodes, which have no direct link, using the hop-by-hop approach and the one-time pad (OTP) encryption.
Building QKD networks without a trusted network requires quantum repeaters~\cite{Zoller1998}, which are now under active research~\cite{Lukin2020,Gisin2020}, but not yet ready for industrial use. 
Thus, the problem of optimizations of the QKD networks deployment is of significant importance~\cite{Lutkenhaus2009,Caleffi2017,Su2017,Mosca2018}.

\begin{figure}[t!]\centering
	\includegraphics[width=0.37\textwidth]{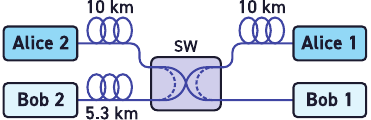}
	\put(-210,60){(a)}
	\\ \vspace{5mm}
	\includegraphics[width=0.4\textwidth]{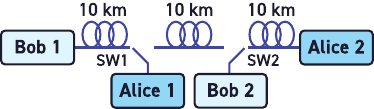}
	\put(-210,60){(b)}
	\caption{\footnotesize Topologies of QKD networks with corresponding optical link lengths between the nodes and the switch, which is controlled by managing Bob\,1. 
	In (a) the star topology is shown. In (b) the backbone topology is illustrated.}
	\label{fig:topology}
\end{figure}

One of the ways for overcoming existing challenges in the QKD network development is to use optical switches, which have been tested in several QKD networks~\cite{Toliver2003,Tang2006}. 
Optical switches allow sharing the resources of the devices in the network, which makes it possible to reduce the number of devices.
An idea behind using switches is that the optical channels of the existing telecommunication structure are very heterogeneous in terms of losses, so the key generation rate varies significantly in different segments.
Therefore, at least in the case of the backbone quantum network configuration, it makes no sense to organize the continuous key generation in all sections -- the key generation rate is limited by the slowest link.
Thus, the use of optical switches in low-loss segments may help to reduce the overall cost of a quantum network significantly, while the secret key rate generation remains reasonably high.
The use of optical switches in QKD networks of various topologies requires further optimization from the viewpoint of time, which is supposed to be fixed for a session between various parties of the network. 
Therefore, an optimization problem on the minimization of the cost of the deployment of the network without a significant reduction of the key generation rate appears to be crucial. 

In this work, we analyze an architecture of QKD networks, which are based on using optical switches and reducing the number of QKD receiver devices, which use single-photon detectors.
We describe the corresponding modification of the QKD network protocol.
We use a realistic model of the performance of QKD devices in realistic experimental conditions.
We present a general framework for optimizing switch-based backbone QKD networks.
As an example, we provide estimations for a network link between of a total of 670\,km length consisting of 8 nodes, 
and show that the switch-based architecture allows achieving significant resource savings of up to 28\%, while the throughput is reduced by 8\% only.

\begin{figure}[]\centering
	\includegraphics[height=0.5\textheight]{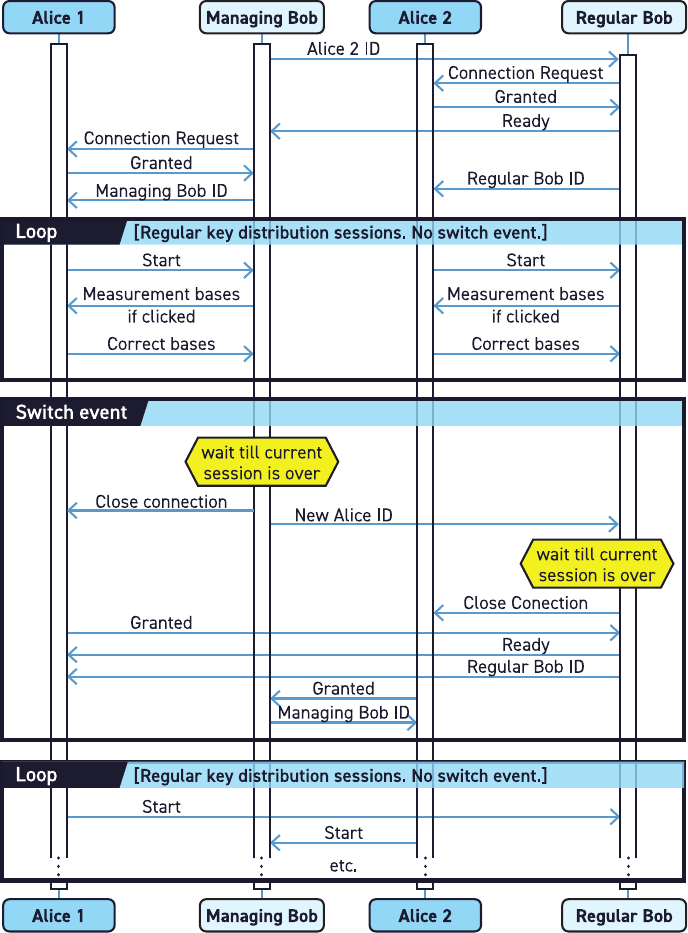}
	\caption{\footnotesize Timing diagram of the QKD realization between four nodes in the star--topology network.}
	\label{fig:block-scheme_2}
\end{figure}

Our work is organized as follows.
In Sec.~\ref{sec:switch}, we describe a switch-based QKD setup.
We provide an analysis of its design and present a model for calculating key rates, which is in good agreement with experimental data. 
In Sec.~\ref{sec:optimization}, we consider a general approach for configuration optimization of switch-based backbone QKD networks.
In Sec.~\ref{sec:Landau}, we use our model of the QKD network to optimize the performance of the backbone QKD network between Moscow and Udomlya.
We summarize the main results of our work and conclude in Sec.~\ref{sec:conclusion}.

\section{Switch-based QKD networks}\label{sec:switch}

Here we briefly describe the extraction of the parameters of the QKD setup, which are further needed for the optimization of the network deployment. 
For this purpose, we develop a model of the performance of realistic QKD devices and compare the predictions of the model with experimental data.
We then apply the developed model for estimating and optimizing the performance of a realistic QKD network using a standard approach for the deployment of (switch-free) backbone QKD networks as a reference.

\subsection{Switch-based quantum networks}

We use optical switches for deploying QKD networks on the basis of QKD setups, which operate using the plug\&play scheme of the BB84 protocol (without decoy states; the scheme is described in Appendix~\ref{sec:app1}).
As usual, we refer to devices where preparation and measurement of qubits is performed as Alice and Bob modules (or Alice and Bob, for short), correspondingly. 

We test two elementary configurations of QKD networks: Star and backbone topologies (see Fig.~\ref{fig:topology}).
In the first case, all the optical channels are loaded continuously, the optical switch is used to commute pairs, for example, Alice\,1--Bob\,1 and Alice\,1--Bob\,2 (see Fig.~\ref{fig:topology}). 
This configuration is suitable for the distribution of security keys, for example, in a branched urban infrastructure. 
Alternative topology, so-called the backbone topology, typically serves to increase the distance of QKD.
For the case of the backbone topology optical switches can be used for optimizing the resources of the network. 
Both elementary configurations might be conveniently interconnected in the necessary order, which ensures the scalability.

The use of QKD devices with optical switched requires modifications of the QKD network protocol.
To control the switches, we need a command from a single point. Therefore, when developing a protocol for the operation of a quantum network, 
we selected a ``managing'' Bob; all other nodes are ``regular'' Bobs (which we refer to as Bobs) and functionally indistinguishable Alices. 
We use standard commercially available optical switch Thorlabs-OSW22-2x2, which is controlled under the host level of the LabVIEW software (which is used for controlling QKD devices, see Appendix~ \ref{sec:app1}). 
The timing diagram of the QKD realization between 4 nodes in the QKD network is shown in Fig.~\ref{fig:block-scheme_2}. 
The exchange is controlled by the ``managing'' Bob\,1, which sends required information via TCP/IP to the ``regular'' Bob\,2.
Bob uses optical switches and adjustment parameters (time windows, train period) and requests the corresponding Alice. 
Alice uses the optical switch, if present, and confirms that the connection is ready. 
For the connection with each Bob, Alice accumulates the generated key in a separate file.
Alice also needs to change the attenuation value of the output signal, depending on the losses in the corresponding quantum channel. 
We note that Alice does not need to change other parameters: The value of the electric delay of the signal from the synchronization detector depends only on the length of the optical storage line of Alice. 
Further, the QKD sessions occur according to the scheme described above until Bob\,1 switch the interacting nodes. 
Obviously, such a scheme can be easily scalable to more interacting nodes.
We note that the commands in the network protocol can be additionally authenticated (or encrypted). 

\subsection{Adjusting a model for QKD devices performance}

\begin{table}[t!]
\begin{center}
	\begin{tabular}{|c|c|c|c|c|}
		\hline
		\multirow{2}{*}{link} & \multicolumn{2}{c|}{theory} & \multicolumn{2}{c|}{experiment} \\
		\cline{2-5}
		& $R_{\rm sift}$\,[kbit/s] & QBER\,[$\%$] & $R_{\rm sift}$\,[kbit/s] & QBER\,[$\%$] \\
		\hline
		Alice\,1--Bob\,1 & 2.2 & 2.7 & 1.8 & 2.6 \\
		\hline
		Alice\,2--Bob\,2 & 2.7 & 0.8 & 2.3 & 1.2 \\
		\hline
	\end{tabular}
	\caption{\footnotesize Comparison of the measured sifted key rate and QBER with theoretical predictions.}
	\label{tab:R-QBER}
\end{center}
\end{table}

\begin{figure}[t!]\centering
	\includegraphics[width=0.45\textwidth]{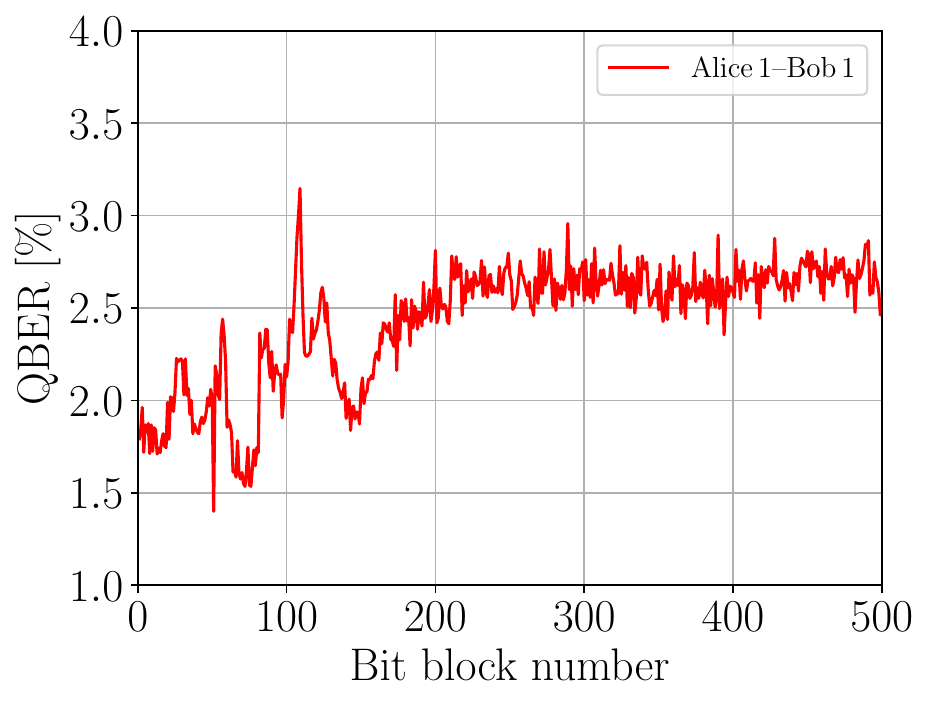}
	\caption{\footnotesize Measured QBER fluctuations in time. The length of each block of the sifted key, for which QBER is computed, is set to be 5000 bytes.}
	\label{fig:QBER_exp}
\end{figure}

In Tab.~\ref{tab:R-QBER}, 
we present our results for experimental measurement of the sifted key rate and quantum bit error rate (QBER) together with corresponding theoretical predictions given by Eqs. from Appendix~\ref{sec:simulation} for the star topology
[see Eqs.~\eqref{eq:R_sift} and \eqref{eq:QBER}]. 
To test and control the measured QBER in time, we split the accumulated sifted key into blocks of length 5000 bytes each, and compute QBER for each block.
Typical QBER fluctuations, which arise mainly due to the imperfect environment of the experiment, are shown in Fig.~\ref{fig:QBER_exp}.

\begin{table*}
\begin{center} 
	\begin{tabular}{| l | l | l |} \hline
		& Alice\,1--Bob\,1 & Alice\,2--Bob\,2 \\ \hline
		Total link length $\ell$\,[km] & 10.0 & 15.3 \\ \hline
		Pulse power at Bob's output $P_B$\,[W] & 1.40$\times10^{-6}$  & 0.13$\times10^{-6}$  \\ \hline
		Total losses in the optical channel $\alpha_{\rm opt}$\,[dB]  & 2.0 & 4.0 \\ \hline
		Internal losses at Alice module $\alpha_A$\,[dB] & 31.7 & 26.5 \\ \hline
		Internal losses at Bob module $\alpha_B$\,[dB] & 7.5 & 6.1 \\ \hline
		Detection efficiency  $\eta_{\rm det}$\,[\%]  &  6& 15 \\ \hline
		Dark count rate DCR\,[Hz] &470  & 110 \\ \hline
		Dead time $\tau_d$\,[$\mu$s] &25 & 10 \\ \hline
		Visibility [\%] & 99.3\,\% & 99.3\,\% \\ \hline
		Afterpulse probability [\%] & 0.1\,\% & 0.1\,\% \\ \hline
	\end{tabular}
	\caption{\footnotesize Measured parameters of links.}
	\label{tab:pars_exp}
\end{center}
\end{table*}

Despite realistic (non-ideal) conditions, after some time since turning on the observed QBER is stabilized, 
and the fluctuations normally do not exceed 3\% for the link Alice\,1--Bob\,1. For the line Alice\,2--Bob\,2 which uses SPD with better characteristics, QBER is found to be about 1\%. 
The average values are given in Tab.~\ref{tab:R-QBER}. 
One can see that there is a rather good agreement between the model and experiment. 

Furthermore, in order to improve the generation rate, one can perform parameter optimization for the current setup by maximizing the lower bound on the secret key rate. 
For fixed and measured optical channel parameters and laser power, we have the following parameters to optimize: the mean number of photons $\mu$, detector efficiency $\eta_{\rm det}$, dead time $\tau_d$ and dark count rate (DCR). 
We use a finite set of points $\{\eta_{\rm det},\tau_d,{\rm DCR}\}$, extracted from the experimental data, and perform the maximization of the lower bound on the secret key rate with respect to $\mu$ as follows:
\begin{equation}\label{eq:R_sec_bound-main}
	R_{\rm sec} = R_{\rm sift} \big\{ \kappa_1^{(l)} [1 - H(E_1^{(u)})] - f_{\rm ec} H(E_\mu) \big\},
\end{equation}
where $R_{\rm sift}$ is the sifted key generation rate,  $\kappa_1$ is a fraction of bits in the verified key obtained from single-photon pulses, 
$E_1$ and $E_\mu$ are the single-photon and the overall QBER's respectively, $f_{\rm ec}$ is the error correction efficiency (here we use $f_{\rm ec}=1.15$ \cite{Trushechkin17}), and $H(\cdot)$ is the binary Shannon entropy.
The detailed discussion of Eq.~\eqref{eq:R_sec_bound-main} is presented in Appendix~\ref{sec:processing}.

In Fig.~\ref{fig:R_sec_opt} we illustrate $R_{\rm sec}(\mu)$ for the fixed SPD parameters listed in Tab.~\ref{tab:pars_exp} (in blue) and for the parameters optimized for each $\mu$--value (in green). 
From the plot it is clearly seen that for the same $\mu\simeq0.4$ the gain in rate by approximately factor 1.5 can be achieved with respect to the current experimental setup (red dot) by just finding an optimal set of $\{\eta_{\rm det},\tau_d,{\rm DCR}\}$. 
Including $\mu$ into optimization variable list as well, numerically we find maximum for the following parameters: $\eta_{\rm det}=30\%$, $\tau_d=25\,\mu$s, DCR$\simeq990$\,Hz and $\mu\simeq0.64$. 
The corresponding maximum key rate is $R_{\rm sec}\simeq2.2$\,kbit/s.
Thus, this simple example proves a clear advantage of parameter optimization.

We note that for our benchmark parameters in Tab.~\ref{tab:pars_exp} the realized standard BB84 protocol can be considered secure, i.e. $R_{\rm sec}>0$, 
only for the losses $\alpha_{\rm opt}$ up to about 5(10)\,dB or equivalently for distances up to 25(50)\,km for the link Alice\,1(2)--Bob\,1(2). 
Therefore, for practical usage at distances $\ell\sim100$\,km one needs an implementation of some other QKD protocol, e.g. the commonly used BB84 with decoy states~\cite{Ma05,Trushechkin21}.

\begin{figure}[h]\centering
	\includegraphics[width=0.45\textwidth]{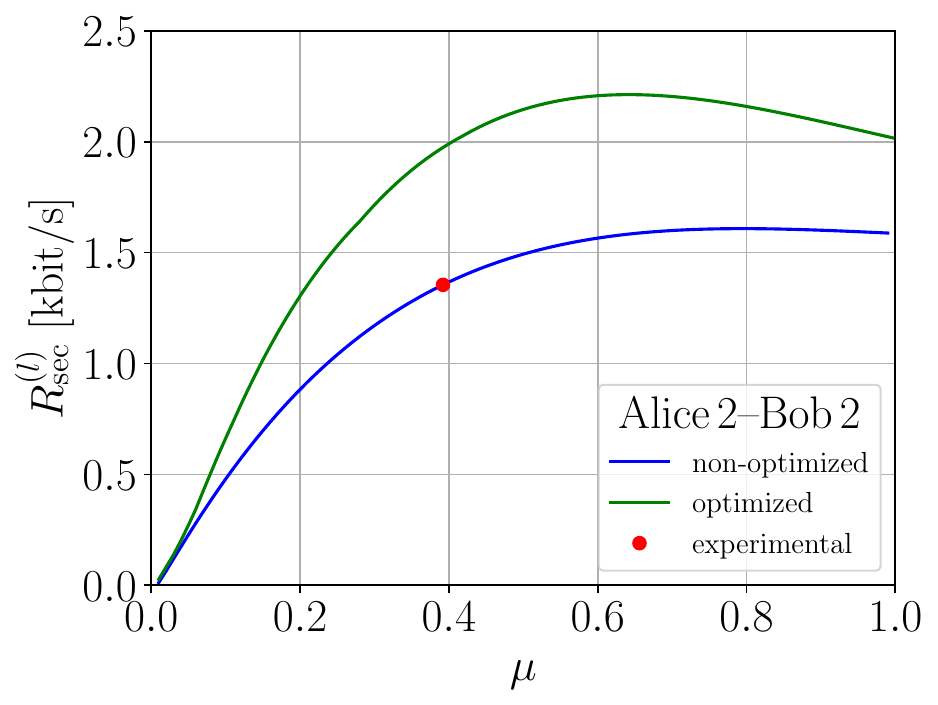}
	\caption{\footnotesize Lower bound on the secret key generation rate for the BB84 protocol without decoy states as function of the mean photon number per pulse with non-optimized (in blue) and optimized (in green) 
	single--photon detector parameters $\{\eta_{\rm det},\tau_d,{\rm DCR}\}$. The red dot represents our current experimental setup with parameters given in Tab.~\ref{tab:pars_exp}.}
	\label{fig:R_sec_opt}
\end{figure}

\section{Optimizing switch-based backbone QKD networks} \label{sec:optimization}

\begin{figure*}
	\centering
	\includegraphics[width=0.75\linewidth]{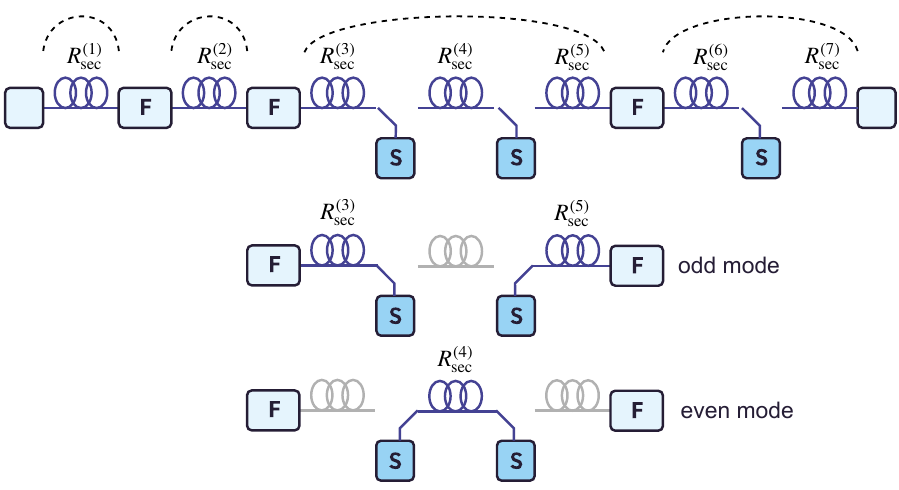}
	\put(-360,190){(a)}
	\put(-275,110){(b)}
	\caption{In (a) an example of a switch-based backbone network with splitting links into clusters is presented. 
	The ``switch-based'' nodes and ``full''nodes with two devices are denoted by S and F, respectively.
	In (b) an operation of a subgroup in odd and even modes is shown.}
	\label{fig:backbone_example}
\end{figure*}

In this section, we describe a general framework for optimizing configurations of switch-based backbone QKD networks.
Consider a sequence of $N$ consecutive QKD links with two edge nodes and $N-1$ trusted nodes.
Let $R^{(i)}_{\rm sec}$ with $i=1,2,\ldots, N$ be secret key generation rates achievable at each link [see Fig.~\ref{fig:backbone_example}(a)].
In the case of no switches in the network, i.e., where all trusted nodes are equipped with two devices (either two Alices, or two Bobs, or Alice and Bob), the resulting key generation rate between the edge node is given by
$\min(R_{\rm sec}^{(1)},\ldots, R_{\rm sec}^{(N)})$, and thus is limited by the slowest link in the network.

Then, let us consider the case where some  trusted nodes are equipped with switches.
We refer these nodes to as ``switch-based'' and denoted them by ${\bf S}$.
In contrast, we call trusted nodes with two devices as ``full'', and denote them by ${\bf F}$.
Thus, the configuration of the network is defined by a string ${\bf X}=({\bf X}_1,{\bf X}_2,\ldots,{\bf X}_{N-1})$, where ${\bf X}_i\in\{{\bf S}, {\bf F}\}$ describes the type of each of $N-1$ trusted nodes.

To calculate a rate for a switch-based backbone network, we split a sequence of all rates $(R_{\rm sec}^{(1)}, \ldots, R_{\rm sec}^{(N)})$ into $m$ clusters
\begin{multline}
	    (R_{\rm sec}^{(1)}, \ldots, R_{\rm sec}^{(N_{1})}), (R_{\rm sec}^{(N_{1}+1)}, \ldots R_{\rm sec}^{(N_{2})}), \ldots,\\ (R_{\rm sec}^{(N_{m}+1)}),\ldots, R_{\rm sec}^{(N)}))
\end{multline}
in such a way that $0<N_1<N_2<\ldots N_m<N$, all trusted nodes connecting links in each of clusters are switch-based, 
and all trusted nodes connecting links between adjacent clusters are full [see an example of splitting links into clusters in Fig.~\ref{fig:backbone_example}(a)].
The basic idea behind the splitting is that since clusters are separated by full trusted nodes, switch-based nodes inside each cluster are able to operate with switches in an independent manner.
The resulting secret key generation between edge nodes in the network then can can be obtained as minimal key generation among all clusters.

Consider a cluster of links with rates $(R_{\rm sec}^{(i_{1})},\ldots,R_{\rm sec}^{(i_{2})})$.
Since all the intermediate nodes within the cluster are switch-based, to achieve the maximal key generation rate through the whole cluster, at each time moment, it is preferable to turn on either all odd or all even links 
[see Fig.~\ref{fig:backbone_example}b].
Assuming that time for switching in negligible, the achievable rate is as follows:
\begin{equation}
	\begin{aligned}
    		&\min(p R_{\rm odd}, (1-p) R_{\rm even}),\\ &R_{\rm odd}=\min\limits_{\substack{i_{1}\leq i \leq i_{2}\\i:{\rm odd}}}R_{\rm sec}^{(i)},\\ &
	    R_{\rm even}=\min\limits_{\substack{i_{1}\leq i \leq i_{2}\\i:{\rm even}}}R_{\rm sec}^{(i)},
    \end{aligned}
\end{equation}
where $p\in(0,1)$ is a parameter determining 
percentage of time in ``even'' and ``odd'' modes.
One can see that the maximal possible rate is given by 
\begin{equation}\label{eq:subgroup}
    R_{\rm subgroup}=\frac{R_{\rm odd}R_{\rm even}}{R_{\rm odd}+R_{\rm even}},
\end{equation}
achieved for $p$ providing $R_{\rm odd}=R_{\rm even}$.

The resulting (effective) key generation rate in the network takes the followng form:
\begin{equation} \label{eq:backbone-rate}
    R_{\rm eff}({\bf X})=\min\left(\left\{R_{\rm subgroup}^{(i)}\right\}_{i=1}^m\right),
\end{equation}
where $R_{\rm subgroup}^{(i)}$ is maximal rate of the $i$th subgroup calculated via Eq.~\eqref{eq:subgroup}, and the splitting of rates into subgroups is determined by the configuration ${\bf X}$.

Using Eq.~\eqref{eq:backbone-rate} one can consider finding the best configuration with a given number of switch-based nodes.
Since in practical scenarios, the total number of nodes usually does not exceed several dozens, this problem can be solved by an exhaustive search approach by trying all possible configurations $\bf X$ and taking a maximum over of all $R(\bf X)$.

The next problem one can consider is an optimization over the cost of equipment related to a given configuration ${\bf X}$.
Let us introduce the following notations: let ${\bf A} ({\bf B})$ denote a node with a single Alice (Bob) device, ${\bf SA} ({\bf SB})$ denote a switch-based node with Alice (Bob) device, 
and ${\bf AA}$, ${\bf AB}$, ${\bf BA}$, ${\bf BB}$ denote a full node with two devices.
One can see that the configuration the same configuration ${\bf X}$ can be implemented in several ways: e.g. ${\bf X}=({\bf S},{\bf F})$ can be realized as {\bf A}-{\bf SB}-{\bf AB}-{\bf A}, {\bf B}-{\bf SA}-{\bf BA}-{\bf B}, {\bf B}-{\bf SA}-{\bf BB}-{\bf A}, and so on.
Though all implementations provide the same key generation rate, their cost can be different because of the different number of Bob and Alice modules in the network (the cost of Alice and Bob module be very different in practice).

In Ref.~\cite{Python-code} we provide an optimization script (upon the reasonable request), 
which allows one to obtain the cheapest implementation of a backbone network with a maximal key generation rate given a sequence of rates, number of switch-based nodes, 
and a cost of each type of node implementation (first, a configuration(s) ${\bf X}$ providing the maximal rate are obtained, and then the cheapest implementation(s) are identified).

\section{Optimization of the QKD network deployment for a realistic network}\label{sec:Landau}

Here we incorporate our results on the analysis and simulation of QKD devices with switches to optimize QKD network deployments.
Below we consider an example of a QKD network (a network link between Moscow and Udomlya).
To estimate the order of magnitude of potential key rates for each link of the network, we use our model for the simulation of QKD devices based on the one-way BB84 with two decoy states. 

For simplicity, we assume that all Bob modules in the network have identical pairs of SPDs with parameters listed in Tab.~\ref{tab:pars_prom}. 
We choose these parameters as reference ones since they describe our currently used experimental decoy-state QKD setup and are studied in detail. 
Although we expect further progress in the impairment of SPD parameters, we use state-of-the-art detectors' parameters in this work. 

By setting the typical order of magnitude values for the signal and decoy states intensities $\mu=0.5$, $\nu_1=0.1$ and $\nu_2=0.01$ respectively, 
we present in Tab.~\ref{tab:Landau} the sifted/secret key rate and QBER estimations for each link of the network (for details of $R_{\rm sec}^{(i)}$ estimation see Appendix~\ref{sec:processing}). 
We note that optimization of $\mu$, $\nu_1$ and $\nu_2$ depending on the channel length is beyond this work, and therefore we assume that the intensities are the same for each link.
One can see that the channel Gorodische--Torzhok has the lowest rate and turns out to be a bottleneck of the network. 
For this reason, one can already conclude that it is unpreferable to put switches at Gorodische and Torzhok nodes.

\begin{table*}[htp] 
\begin{center}
	\begin{tabular}{|l|c|}
		\hline
		Pulse repetition frequency $f$ & 312.5\,MHz \\ \hline
		Gate time window $\tau_{\rm gate}$& 600\,ps \\ \hline
		Internal losses at Bob module $\alpha_B$ & 4\,dB \\ \hline
		Detection efficiency $\eta_{\rm det}$ & 10\,\% \\ \hline
		Dark count rate DCR & 300\,Hz \\ \hline
		Dead time $\tau_d$ & 5\,$\mu$s \\ \hline
		Visibility $V$ & 98\,\% \\ \hline
		Afterpulse probability $p_{\rm after}$ & 3\,\%\\
		\hline
	\end{tabular}
	\caption{Parameters of single photon detectors used for modeling QKD devices.}
	\label{tab:pars_prom}
\end{center}
\end{table*}

\begin{table*}[htp] 
\begin{center}
	\begin{tabular}{|c|c|c|c|c|c|}
		\hline
		Link & Length [km] & Optical losses [dB] & Sifted key rate\,[kbit/s] &Secret key rate ($R_{\rm sec}^{(i)}$)\,[kbit/s] & QBER\,[\%] \\
		\hline
		Moscow--Kubinka			& 86.8  & 19.0 & 16.8 & 2.7 & 4.1 \\
		\hline
		Kubinka--Uvarovka		& 115.0 & 22.2 & 8.8  & 1.4 & 4.2 \\
		\hline
		Uvarovka--Gagarin		& 74.0  & 14.6 & 35.7 & 5.9 & 4.0 \\
		\hline
		Gagarin--P.\,Gorodische & 98.7  & 18.9 & 17.1 & 2.8 & 4.1 \\
		\hline
		P.\,Gorodische--Torzhok & 125.8 & 23.6 & 6.6  & 1.0 & 4.2 \\
		\hline
		Torzhok--V.\,Volochek   & 114.4 & 21.8 & 9.6  & 1.5 & 4.1 \\
		\hline
		V.\,Volochek--Udomlya   & 82.5  & 15.9 & 29.2 & 4.8 & 4.0 \\
		\hline
	\end{tabular}
	\caption{Model parameters for the backbone QKD network between Moscow and Udomlya.
		BB84 protocol with signal intensity $\mu=0.5$, and two decoy states for intensities $\nu_1=0.1$ and $\nu_2=0.01$ is considered. 
		Single photon detector parameters are taken from Tab.~\ref{tab:pars_prom}.}
	\label{tab:Landau}
\end{center}
\end{table*}

\begin{figure*}[t!]\centering
	\includegraphics[width=0.6\textwidth]{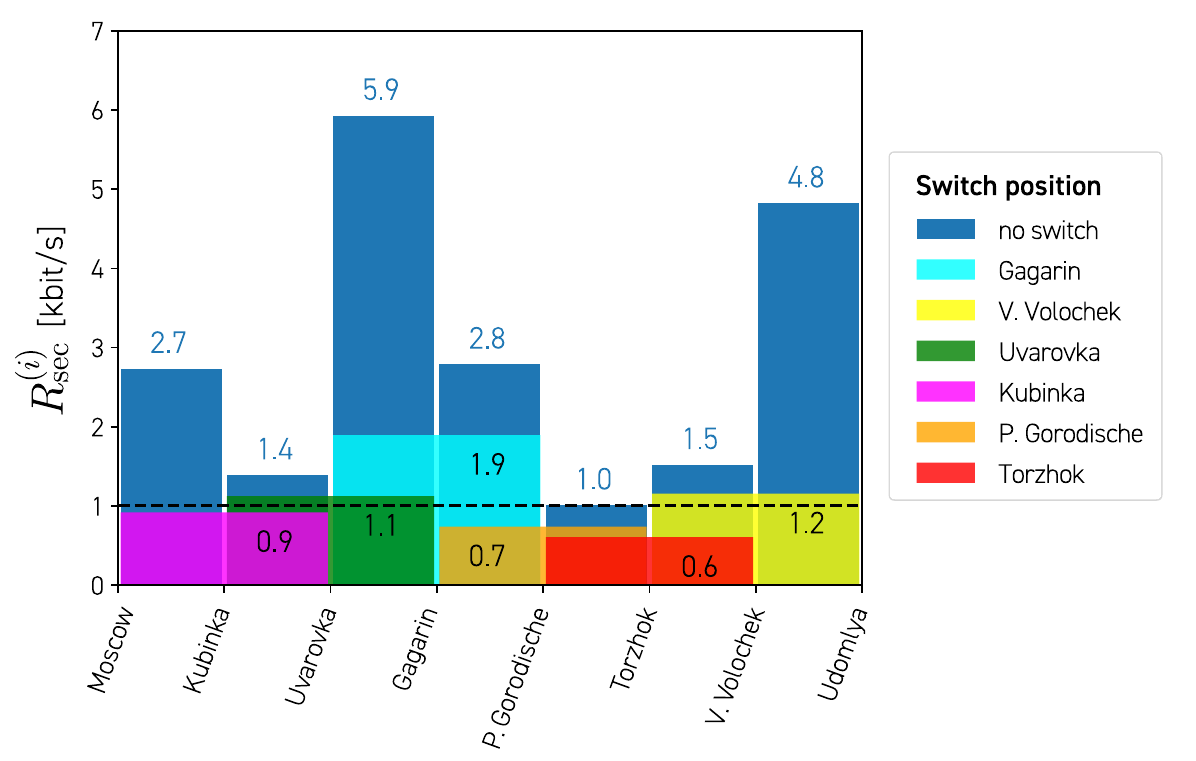}
	\caption{\footnotesize Secret key generation rates from Tab.~\ref{tab:Landau} for each link (in blue). 
	Wide coloured bars represent the rate of adjacent links with an intermediate switch. 
	The black dashed line marks the minimum rate in the network.}
	\label{fig:R_Landau_chart}
\end{figure*}

In order to demonstrate how the key generation rates are affected by introducing switches in the network, we first consider the case of a single switch.
According to Eq.~\eqref{eq:subgroup}, introducing a switc is equivalent to replacing a pair of adjacent links with secret key generation rates 
$R_{\rm sec}^{(i)}$ and $R_{\rm sec}^{(i+1)}$ with a single effective link with rate $R_{\rm sec}^{(i)}R_{\rm sec}^{(i+1)}/(R_{\rm sec}^{(i)}+R_{\rm sec}^{(i+1)})$.
In Fig.~\ref{fig:R_Landau_chart}, we present the comparison of original key generation rate of all links and effective key generation rates appeared after introducing switches at different nodes.
Since the resulting rate between edge nodes is given by minimum of all rates, putting switches in Uvarovka, Gagarin, or V. Volochek does not decrease the rate, though reduce the number of  employed devices.

It turns out that the resulting effective key rate of 1\,kbit/s does not decreases even in the case of putting switches at all three mentioned nodes (see Tab.~\ref{tab:Landau_2}).
With three switches, the total number of Alice and Bob modules equals to 11, while for the classic backbone line 14 modules are required. 
This trivial comparison demonstrates a clear economic benefit of the proposed switch line: If Alice and Bob devices had the same cost, na\"{i}vely the benefit would be 21\%.
We also note that the configuration of the network with switches in Uvarovka, Gagarin, and V. Volochek can be realized with different number of Alice and Bob modules.
The configurations  
\begin{equation}
	\begin{aligned}
		&{\bf A}-{\bf BA}-{\bf SB}-{\bf SA}-{\bf BA}-{\bf BA}-{\bf SB}-{\bf A},\\
		&{\bf A}-{\bf BA}-{\bf SB}-{\bf SA}-{\bf BA}-{\bf BB}-{\bf SA}-{\bf B} 
	\end{aligned}
\end{equation}
have the same number of switches and provide the same rate, however the first configuration has less number of Bob modules than the second one.
Taking into account that Bob module contains more complex hardware components, in particular, SPD, that is not only a very important and sophisticated but also a rather sensitive and expensive device, 
having less Bob modules in the network apparently reduces the maintenance cost.

Further increase in the number of switches leads to a decreasing of the effective key generation rate as well as the total number of modules and corresponding equipment cost (see Tab.~\ref{tab:Landau_2}).
Note that using four switches compared to the case of no switches drops down the number of modules from 14 to 10 ($\approx 28\,\%$ decrease), while decreases the effective key generation rate by 8\,\% only.
Putting switches in all nodes results in decrease of the key generation rate down to 0.58\,kbit/s and reducing the number of employed modules down to 8 units.
Thus, we observe a clear tradeoff between the effective key generation rate and the implementation cost.
The choice of switches number can be made based on requirements posed on the network by the surrounding infrastructure.

\begin{table*}[htp] 
\begin{center}
	\begin{tabular}{|c|c|c|c|c|c|}
		\hline
		& $R_{\rm eff}$\,[kbit/s] & $N_A$ & $N_B$ & $N_{\rm tot}$ & Variant of implementation\\
		\hline
		without switches	(backbone)		   & 1.0 & 7 & 7 & 14 & {\bf A}-{\bf BA}-{\bf BA}-{\bf BA}-{\bf BA}-{\bf BA}-{\bf BA}-{\bf B}\\
		\hline
		switches in Uvarovka, Gagarin \& V.\,Volochek & 1.0 & 6 & 5 & 11 & {\bf A}-{\bf BA}-{\bf SB}-{\bf SA}-{\bf BA}-{\bf BA}-{\bf SB}-{\bf A}\\
		\hline
		+switch in Kubinka					   & 0.92 & 6 & 4 & 10 & {\bf A}-{\bf SB}-{\bf SA}-{\bf SB}-{\bf AA}-{\bf BA}-{\bf SB}-{\bf A} \\
		\hline
		+switch in P.\,Gorodische				   & 0.6 & 5 & 4 & 9 & {\bf A}-{\bf SB}-{\bf SA}-{\bf SB}-{\bf AB}-{\bf SA}-{\bf SB}-{\bf A} \\
		\hline
		+switch in Torzhok					   & 0.58 & 4 & 4 & 8 & {\bf A}-{\bf SB}-{\bf SA}-{\bf SB}-{\bf SA}-{\bf SB}-{\bf SA}-{\bf B}\\
		\hline
	\end{tabular}
	\caption{\footnotesize Effective key rate in the backbone network  $R_{\rm eff}$, required number of Alice ($N_{A}$) and Bob ($N_{B}$) modules, total number of modules ($N_{\rm tot}$), and variants of implementation assuming that Bob module is more expensive than Alice one.}
	\label{tab:Landau_2}
\end{center}
\end{table*}

\section{Conclusion and outlook}\label{sec:conclusion}

In the present work, we investigated QKD network deployment based on intermediate nodes with optical switches. 
For the proof-of-principle, we realized in practice a laboratory BB84 plug\&play scheme QKD network prototype with two Alice and two Bob nodes connected via an optical switch. 
A good agreement between the experimental key rates and our theoretical model estimations was found. 
Hence, we applied our results for making predictions for a QKD network between Moscow and Udomlya, which has 8 nodes and a total length of 670\,km.
We investigated possible line configurations with various Alice/Bob/switch device placing and proposed several configurations that provide the entire system maintenance cost reduction without essential loss in the overall rate. 
In particular, it was demonstrated that one could achieve cost savings of up to 28\,\% while reducing throughput by 8\,\%.

\section*{Acknowledgments}
This work is supported by the Russian Science Foundation under project 17-71-20146.
We thank PJSC Rostelecom for providing data on the network between Moscow and Udomlya.

\appendix

\section{QKD setup}\label{sec:app1}

We use a two-pass auto compensation QKD plug\&play scheme, which is based on the modular QKD platform for research and education~\cite{Fedorov18,Rodimin19}.
It operates on the basis of the original BB84 protocol~\cite{BB84} with or without decoy states~\cite{Ma05,Trushechkin21}.

Each node of the hardware part of the platform (Alice or Bob) consists of a PC with the R--series board PCIe--4820R by National Instruments (NI) on which all control signals are generated. 
The control signals are sent via a stranded digital cable VHDCI to a specially designed motherboard. 
To this motherboard there can be connected up to 12 add--on cards, such as laser modules, phase and amplitude modulator drivers, the module with MEMS attenuator, a fast photodetector for synchronization in the plug\&play scheme.
On the motherboard, we placed the Spartan--6 FPGA chip whose main function is to route signals between the add--on cards and the NI R--series board. 
The software for each node of Alice or Bob consists of the FPGA level and the host level.

The FPGA level is responsible for the laser pulse generation, pulse-train formation and overall signal control, as ones for phase modulators, etc. 
Also at the FPGA level, the single-photon detector (SPD) signals are recorded in the FIFO memory, if the signals arrive within the correct time window. 
In order to simplify the setup and to reduce the cost of the device, we implement the scheme with one SPD in the free-running regime. 
To reduce the errors due to dark counts, the SPD signals are registered within a time window equal to $1/20$ of the laser pulse repetition period, $T=1/f=200$\,ns, where $f=5$\,MHz is the pulse repetition rate. 
The used sets of SPD parameters, as well as fiber lengths, optical losses and laser power, are be presented in Tab.~\ref{tab:pars_exp}.

For the preparation and measurement of quantum states, one needs to use a random number generator (RNG).
In order to guarantee the security of the corresponding quantum-generated keys quantum random number generator should be used. 
The preparation and measurement bases and bits values, recorded in the FIFO FPGA, are transferred to the host software level and then the obtained raw key is sifted and distilled. 

In order to calculate the QKD session time, we take into account only the FPGA level operation time (train generation, measurement of quantum states), 
because sifting and necessary post-processing procedures can be parallelized with the FPGA hardware operation. 
During one session we generate 1000 trains of 1200 laser pulses in each train. 
The train period $T_t$ was made as minimal as possible and ranged from $600-660\,\mu$s, depending on the length of the quantum channel. 
Even for not very powerful computers, for $f=5$\,MHz the operating time of the hardware exceeds the execution time of the host procedures.

\section{Simulation of the QKD devices performance}\label{sec:simulation}

Here we describe a model of QKD devices performance, which is based on the standard plug\&play scheme BB84 protocol but with only one single-photon detector 
(at the time of our experiments only one detector per Alice device was available; it can further be easily generalized for the one-way decoy-state protocol with two detectors).

For continuously sent periodic pulses, the sifted key rate can be simply written as follows:
\begin{equation}\label{eq:R_sig_0}
	R = {1\over4} f p_{\rm sig} =  {1\over4} f (1 - e^{-\eta\mu}) \simeq {1\over4} f \eta\mu,
\end{equation}
where $p_{\rm sig}$ is the probability to emit and receive a signal pulse containing at least one photon, 
$\mu$ is the attenuated intensity of the pulse, going out of Alice back to Bob, and $\eta$ is the overall efficiency of the transmission from Alice to Bob and further detection and is given by the following expression:
\begin{equation}
	\eta = \eta_{\rm det} \times 10^{-0.1(\alpha_{\rm opt} + \alpha_B)},
\end{equation}
with detector's efficiency $\eta_{\rm det}$, total internal losses at Bob $\alpha_B$  and in the optical link $\alpha_{\rm opt}$. 
Only the events when Alice and Bob have compatible bases are considered, therefore only half of the raw key is taken into account. 
We also do not consider errors when incompatible bases are used. Note that for the setup with only one detector Bob has to choose randomly not only the basis, but also the information bit that he plans to check in this basis. This in turn reduces the key rate by factor two compared to the scheme with two detectors.

Due to high repetition rate the effects of Rayleigh backscattering in fiber can be non-negligible. 
Some strong pulses, still going to Alice, can be backscattered on fiber inhomogeneities during the intersection with already returning weakened pulses, and travel together back to Bob. 
Despite rather low backscattering probability, because of very high intensity the backscattered signals can induce false detection counts and thus increase the quantum bit error rate (QBER). 
To solve this problem, it is commonly used in plug\&play schemes to introduce a storage line (SL) in Alice module. Bob emits pulses not continuously but in trains of $N_p=1200$ pulses per train that travel to Alice and fill SL. 
The SL length is adjusted such that $2\ell_{\rm SL}\simeq(N_p-1)Tc/n_{\rm fib}\simeq50$\,km, where $T=1/f$ is the pulse period in the train, and $n_{\rm fib}\simeq1.47$ is the optical fiber refractive index. 
After the last pulse of the train is emitted, Bob waits for the time required to reach Alice and come back, and then sends the next train. 
Thus, the train period is chosen to be $T_t\simeq(N_p-1)T+2(\ell+\ell_{\rm SL})n_{\rm fib}/c\simeq0.6$\,ms for $\ell=10,15$\,km. In this way, the intersection of forward and backward pulses takes place only in SL and not in the transmission line. 
Since SL is located after VOA, the intensity is significantly low, so that the probability of backscattering inside SL to induce false detections can be neglected. 
We note that the signal losses in SL are already taken into account when we estimate the required VOA attenuation in order to obatin the desired $\mu$ and therefore do not affect $\eta$. 
However, the effective repetition frequency and hence the key rate are reduced due to the time gaps while waiting for the previous train to come back before emitting a next one:
\begin{equation}\label{eq:R_sig}
	R = {1\over4} f_{\rm eff}\, \eta \mu = {1\over4} f {N_p \over T_t / T} \eta \mu \simeq {1\over4} f {\ell_{\rm SL} \over \ell + 2\ell_{\rm SL}} \eta \mu .
\end{equation}
One can mention that for the scheme with two SPDs the result in Eq.~\eqref{eq:R_sig} has to be multiplied by factor 2.

The QBER value is defined can be expressed in terms of rates of receiving a false/correct detection per pulse as follows:
\begin{equation}\label{eq:QBER}
\begin{split}
	\QBER&= {\rm false~counts \over total~counts}  \\ 
	&\simeq{p_{\rm dark} + (p_{\rm sig} + p_{\rm dark}) p_{\rm after} + p_{\rm sig} p_{\rm opt} \over p_{\rm sig}} .
\end{split}	
\end{equation}
where $p_{\rm dark}$ is the dark count probability,
$p_{\rm opt}$ is the probability of a photon to hit a wrong detector due to flipped and incorrectly determined polarization or phase, 
and can be related to the visibility as $p_{\rm opt}=(1-V)/2$; $p_{\rm after}$ is the cumulated probability to register an afterpulse since the end of the detector's dead time $\tau_d$, 
which is triggered by a previously detected signal photon or dark count that yielded an avalanche.
We note that with 50\% of chance a dark count will occur when Bob chooses the correct bit value, and hence will not lead to an error but just to an additional count. 
Here we take into account only the events with compatible bases.

In order to reduce the dark count contribution, the detector can be biased above the breakdown voltage during a short period of time, $\tau_{\rm gate}$ which is of order of pulse width. 
This gate is periodically repeated with pulse frequency $f$. 
In this way, the detector is sensitive only during the gates when the signal photons are expected. The dark count probability is then given by the following expression: 
\begin{equation}
	p_{\rm dark} = {\rm DCR} \cdot \tau_{\rm gate} ,
	\label{eq:p_dark}
\end{equation}
where DCR stands for dark count rate in Hz (it is a key parameter of the detector). 
We do not impose the gated detection, and thus our detector works in the free-running mode. 
In this case, we have simply $\tau_{\rm gate}=T$.

\begin{figure}[t!]\centering
	\includegraphics[width=0.45\textwidth]{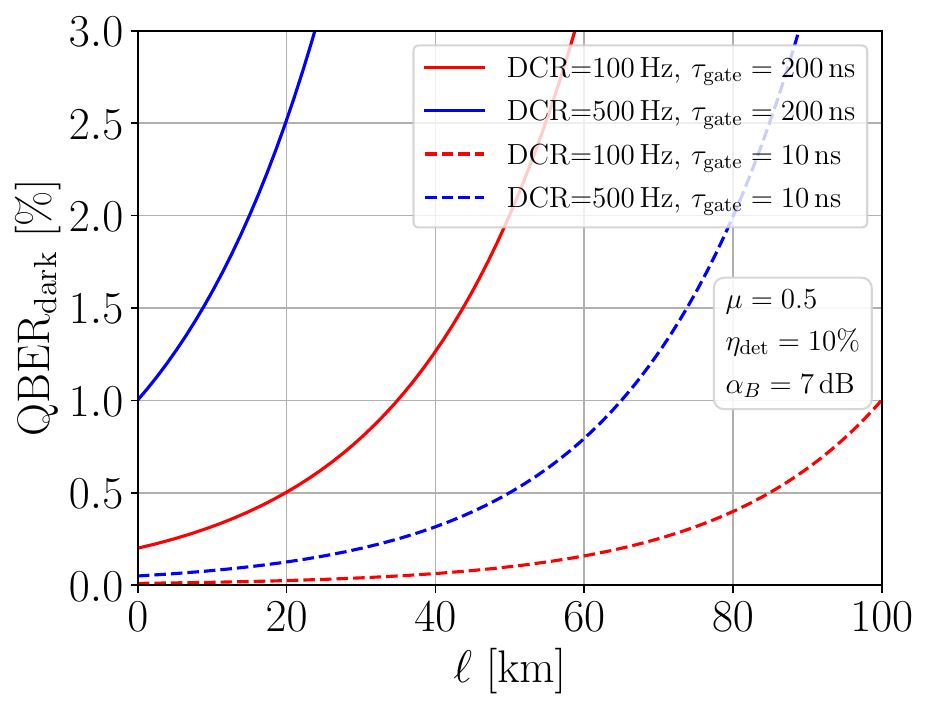}
	\caption{\footnotesize The dark count contribution to QBER as function of the optical link length.}
	\label{fig:QBER_dark}
\end{figure}

For example, for $\mu=0.5$ and $\eta_{\rm det}=10\%$ and the typical probabilities $p_{\rm opt/after}\sim1\%$, 
one can see from Fig.~\ref{fig:QBER_dark} that for distances 10--20\,km the dark count contribution to the total QBER becomes significant or even dominant with respect to the optical or afterpulse contribution only for large DCR and wide $\tau_{\rm gate}$.

We note that DCR and $\eta_{\rm det}$ are not independent parameters. 
DCR increases with $\eta_{\rm det}$, as well as their ratio. In particular, DCR and $\eta_{\rm det}$ both go up with increasing bias voltage $U_{\rm bias}$. 
Therefore, finding an optimal $\eta_{\rm det}$ is a tradeoff between high key rate and low QBER. 

\begin{figure}[t!]\centering
	\includegraphics[width=0.45\textwidth]{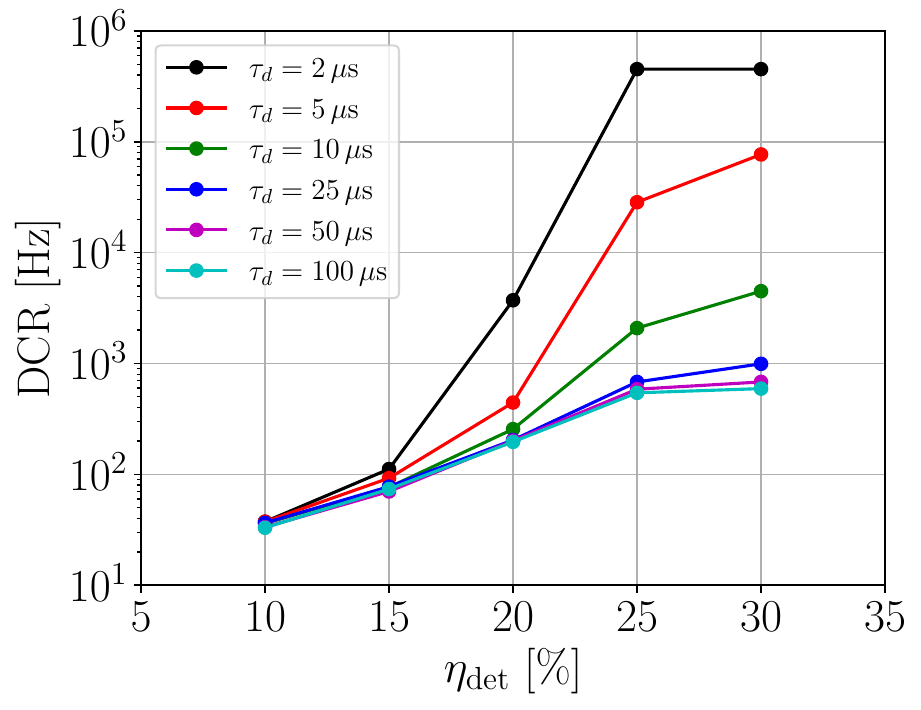}
	\caption{\footnotesize Dark count rate (DCR) as function of the detector's efficiency $\eta_{\rm det}$ of the InGaAs/InP single photon detector ID230.}
	\label{fig:DCR_IDQ}
\end{figure}

Let us estimate the effects of the detector's dead time, $\tau_d$. 
We assume that the real events (both signal, dark counts and afterpulses) that occur during the dead period are lost and have no effect on the system. 
The fraction of all time that the detector is dead is simply the product $R_{\rm sift}\tau_d$, where $R_{\rm sift}$ is the measured sifted key rate. Neglecting the small background contribution, the rate of loss can be written as follows:
\begin{equation}
	R - R_{\rm sift} = R R_{\rm sift}\tau_d.
	\label{eq:dead_time}
\end{equation}
Solution to Eq.~\eqref{eq:dead_time} has the following form:
\begin{equation}
	R_{\rm sift} = {R \over 1 + R \tau_d} .
	\label{eq:R_sift}
\end{equation}
We note that Eq.~\eqref{eq:R_sift} is valid for both plug\&play and one-way schemes. 
Apparently, reducing the dead time increases the detection statistics and thus increases the raw/sifted key rate. 
However, at the same time it increases DCR (see Fig.~\ref{fig:DCR_IDQ}) and consequently QBER, then lowering the final key rate after information reconciliation~\cite{Kiktenko2016,Fedorov2017,Kiktenko20172}. 
Thus, as in the case with $\eta_{\rm det}$, the choice of $\tau_d$ becomes a matter of tradeoff between high statistics and low QBER.

\section{Post--processing}\label{sec:processing}

Here we estimate roughly the final secret key rate that could be obtained with our setup including the post-processing stage~\cite{Kiktenko2016,Fedorov2017,Kiktenko20172}.
In the theoretical limit with infinite number of pulses, the secret key rate is given by
\begin{equation}
	R_{\rm sec} = R_{\rm ver} \big\{ \kappa_1 \big[1 - H(E_1)\big] - f_{\rm ec} H(E_\mu) \big\},
	\label{eq:R_sec}
\end{equation}
where $R_{\rm ver}$ is the verified key rate, i.e. the rate after the information reconciliation, 
$\kappa_1=Q_1/Q_\mu$ is the fraction of bits in the verified key obtained from single-photon pulses, 
$E_1$ and $E_\mu$ are the single-photon and the overall QBER's respectively, $f_{\rm ec}$ is the error correction efficiency (here we use $f_{\rm ec}=1.15$ \cite{Trushechkin17}), and $H$ is the binary Shannon entropy,
\begin{equation}
	H(x) = -x \log_2(x) - (1 - x) \log_2(1 - x).
\end{equation}
We neglect statistical fluctuations and assume that we know the exact values of $\kappa_1$ and $E_1$. As information reconciliation scheme one can choose the error correction with e.g. low-density-parity-codes (LDPC). 
Assuming for simplicity the LDPC frame error rate (FER) to be zero, 
we can use the approximation $R_{\rm ver}\approx R_{\rm sift}$ (in practice we find that ${\rm FER}\sim10^{-4}$ can be achieved, so our assumption is reasonable). 
For the estimation of the single-photon and overall gains $Q_1$ and $Q_\mu$ and respective errors $E_1$ and $E_\mu$; see e.g. Refs.~\cite{Ma05,Trushechkin21}.

In the standard BB84 without decoy states, $Q_1$ and $E_1$ are not directly measured, only $Q_\mu$ and $E_\mu$ are. 
Eve is able to block single photon pulses and use ideal lossless channel to retransmit at most one photon from all multi-photon pulses to Bob. 
Therefore, in order to estimate the worst-case scenario secure key rate we have to assume that all losses and errors are from single-photon pulses.
We also assume that Eve cannot break in Bob's device and control photon detectors.
For this Eve's sub-optimal strategy, we get the lower bound on $Q_1$ (and hence $\kappa_1$) and the upper bound on $E_1$,
\begin{equation}
	\begin{split}
		&Q_1^{(l)} = Q_\mu - \eta_{\rm det} \times 10^{-0.1\alpha_B} \sum_{n=2}^\infty {\mu^n \over n!} e^{-\mu} \\
		&= Q_\mu - \big[1 - (1 + \mu) e^{-\mu} \big] \eta_{\rm det} \times 10^{-0.1\alpha_B} \\ 
		& = Q_\mu \kappa_1^{(l)},
	\end{split}
\end{equation}
\begin{equation}
	E_1^{(u)} = {Q_\mu E_\mu \over Q_1^{(l)}},
\end{equation}
giving the lower bound on the key rate,
\begin{equation}
	R_{\rm sec} = R_{\rm sift} \big\{ \kappa_1^{(l)} [1 - H(E_1^{(u)})] - f_{\rm ec} H(E_\mu) \big\}.
	\label{eq:R_sec_bound}
\end{equation}

Now let us consider the decoy-state BB84 protocol. 
Following Refs.~\cite{Ma05,Trushechkin21}, let $\mu$, $\nu_1$ and $\nu_2$ be the signal, weak decoy and vacuum decoy state intensities respectively. 
Then the lower bounds on the background and one--photon yields can be estimated as follows:
\begin{equation}
	Y_0^{(l)} = {\rm max} \bigg\{ {\nu_1 Q_{\nu_2} e^{\nu_2} - \nu_2 Q_{\nu_1} e^{\nu_1} \over \nu_1 - \nu_2}\,, 0 \bigg\},
	\label{eq:Y_0_decoy}
\end{equation}
\begin{equation}\label{eq:Y_1_decoy}
\begin{split}
	&Y_1^{(l)} = {\mu \over (\nu_1 - \nu_2) (\mu - \nu_1 - \nu_2)} \\
	&\times\bigg[ Q_{\nu_1} e^{\nu_1} - Q_{\nu_2} e^{\nu_2} - {\nu_1^2 - \nu_2^2 \over \mu^2} (Q_\mu e^\mu - Y_0^{(l)}) \bigg].
\end{split}
\end{equation}
The upper bound on one-photon QBER is given by the following expression:
\begin{equation}
	E_1^{(u)} = {E_{\nu_1} Q_{\nu_1} e^{\nu_1} - E_{\nu_2} Q_{\nu_2} e^{\nu_2} \over (\nu_1 - \nu_2) Y_1^{(l)}}.
	\label{eq:E_1_decoy}
\end{equation}
The bound on the key rate is computed from Eq.~\eqref{eq:R_sec_bound} using $\kappa_1^{(l)}=Y_1^{(l)}\mu e^{-\mu}/Q_\mu$ and Eqs.~\eqref{eq:Y_0_decoy}--\eqref{eq:E_1_decoy}. 
We note that since the state type is chosen randomly with corresponding probability $p_{\mu/\nu_1/\nu_2}$ ($p_{\nu_2}=1-p_\mu-p_{\nu_1}$), 
Eq.~\eqref{eq:R_sig} must be multiplied by $p_\mu$ when computing $R_{\rm sift}$ \eqref{eq:R_sift}. In this work we set $p_\mu=0.5$ and $p_{\nu_1}=p_{\nu_2}=0.25$.

\end{document}